\documentclass[12pt]{article}

\usepackage{amsmath,amssymb}
\usepackage{graphicx}

\makeatletter

\@addtoreset{equation}{section}
\makeatother

\newcommand{\be}{\begin{equation}}
\newcommand{\ee}{\end{equation}}
\newcommand{\bea}{\begin{eqnarray}}
\newcommand{\eea}{\end{eqnarray}}
\newcommand{\beann}{\begin{eqnarray*}}
\newcommand{\eeann}{\end{eqnarray*}}

\newcommand{\ba}{\begin{array}}
\newcommand{\ea}{\end{array}}

\newcommand{\Tr}{\mathop{\rm Tr}}

\newcommand{\R}{\mathbb{R}}
\newcommand{\C}{\mathbb{C}}
\newcommand{\G}{\mathbb{G}}
\newcommand{\Z}{\mathbb{Z}}
\newcommand{\p}{\partial}
\newcommand{\D}{{\cal D}}
\newcommand{\tQ}{\tilde{Q}}
\newcommand{\tq}{\tilde{q}}
\newcommand{\M}{{\cal M}}
\newcommand{\E}{{\cal E}}
\newcommand{\cG}{{\cal G}}
\newcommand{\Q}{{\cal Q}}
\newcommand{\cP}{{\cal P}}
\newcommand{\W}{{\cal W}}
\newcommand{\N}{{\cal N}}

\def\XXint#1#2#3{{\setbox0=\hbox{$#1{#2#3}{\int}$} 
\vcenter{\hbox{$#2#3$}}\kern-.5\wd0}}

\begin{document}

\setlength{\oddsidemargin}{0cm}
\setlength{\baselineskip}{7mm}

\begin{titlepage}
\renewcommand{\thefootnote}{\fnsymbol{footnote}}

RIKEN-TH-173

\vspace*{0cm}
    \begin{Large}
    \begin{bf}
       \begin{center}
         { New Gauged Linear Sigma Models for\\
8D HyperK{\"a}hler Manifolds\\ and \\Calabi-Yau Crystals}     
       \end{center}
    \end{bf}   
    \end{Large}
\vspace{1cm}

   \begin{large}
\begin{center}
 { \sc Yutaka Baba}\footnote    
{e-mail address : 
ybaba@riken.jp} and {\sc Ta-Sheng Tai}\footnote    
{e-mail address : 
tasheng@riken.jp} 
\end{center}
\end{large}

\begin{center}
{\it   Theoretical Physics Laboratory, RIKEN,
                    Wako, Saitama 351-0198, JAPAN}
\end{center}

\begin{abstract}
\noindent
We propose two kinds of gauged linear sigma models whose moduli spaces are real eight-dimensional hyperK{\"a}hler and Calabi-Yau manifolds, respectively. 
Here, 
hyperK{\"a}hler manifolds 
have $sp(2)$ holonomy in general and are 
dual to Type IIB $(p,q)$5-brane configurations. 

On the other hand, Calabi-Yau fourfolds are 
toric varieties expressed as quotient spaces. 
Our model involving fourfolds is different from the usual 
one which is directly 
related to a symplectic quotient procedure. Remarkable 
features in newly-found three-dimensional Chern-Simons-matter 
theories appear here as well, such as $dynamical$ Fayet-Iliopoulos 
parameters, one 
$dualized$ $photon$ and its 
residual discrete gauge symmetry. 
\end{abstract}
\vfill 

\end{titlepage}
\vfil\eject

\setcounter{footnote}{0}

\section{Introduction}
A gauged linear sigma model (GLSM) in two dimensions 
\cite{Witten:1993yc} is capable of describing a curved 
geometry (typically a sympletic or K{\"a}hler quotient space) 
through gauge theory language. 
More precisely, the isometry%
\footnote{For a K{\"a}hler quotient space $M//G$, 
the isometry $G$ is required to be 
sympletic such that 
vector fields $K$'s generating it give 
$ {\cal L}_{K} \omega=0$ where 
$\omega$ is 
a sympletic form on the 
sympletic manifold $M$. } 
of the transverse 
internal space gets partially gauged and 
chiral multiplets 
on the worldsheet are coupled to corresponding gauge fields. 
Then, the curved geometry 
$\C^n//G$ arises 
as the supersymmetric vacuum 
(moduli space) 
of the 2D gauge theory. 
Note that 
$G=\C^{\star r}$ stands for $r$ complexified $U(1)$'s 
with moment maps
\begin{eqnarray}\label{DM1}
\mu: ~\C^n \to \R^{r}, ~~~~~\mu^a=
\sum^n_{i=1} Q^a_i ~|\phi_i|^2, ~~~~~a=1, \dots, r, ~~~~~
Q^a_i: \mathrm{charge ~assignment}.
\end{eqnarray}
$\phi_i$ parameterizing 
$\C^n$ represents the lowest component 
of an $\N$=(2,2) chiral superfield 
$\Phi_i$. Also, $\mu^a$'s are called 
Fayet-Iliopoulos (FI) parameters which bring 
in $r$ K{\"a}hler classes. \eqref{DM1} modulo the following 
gauge symmetry 
\begin{eqnarray}\label{}
\phi_i \sim \phi_i ~e^{i\sum_a Q^a_i \lambda^a},  ~~~~~~\lambda^a \in \mathbb{R} 
\end{eqnarray}
is exactly the vacuum manifold denoted as 
$\C^n//G$ or ${\mu}^{-1} (0)/U(1)^r$.

Similarly, 
a hyperK{\"a}hler quotient space is defined by 
$\mathbb{H}^n///\G$ 
($\mathbb{H}^n\cong\mathbb{R}^{4n}$) where 
$ \G (\subset {\mathrm{ triholomorphic ~
isometry}} )$ 
is generated by vector fields $K$'s 
with ${\cal L}_{K} g=0$ ($g$: metric) and ${\cal L}_{K} I^\alpha=0$. 
Three complex structures $I^\alpha$ 
$(\alpha=1,2,3)$ are  
\begin{eqnarray}\label{}
\omega^\alpha = 
d\cP^\dagger  { \sigma}^\alpha
\wedge d\cP, ~~~~\cP=(q, \tilde q^\dagger)^t,  ~~~~
{ \sigma}^\alpha: \mathrm{Pauli~ matrices}
\end{eqnarray}
for each $\mathbb{H}$ described by 
$ds^2= |dq|^2 + |d \tq|^2$ 
$(q, \tilde q\in \mathbb{C})$ and transform as a triplet 
under $SU(2)$. Our convention is as follows. 
A quaternion $\Q=y + {\boldsymbol y}~ (\bar{\Q}=y-{\boldsymbol y} )$ with a pure imaginary part ${\boldsymbol y}=ui + vj +wk$ 
$(y,u,v,w \in {\mathbb{R}})$ 
consists of 
a pair of complex numbers $(q, \tilde q)$: 
\begin{eqnarray}\label{q}
\Q=-q + \tilde q j, ~~~~~~
\bar{\Q}=-q^\dagger -j \tq^\dagger.
\end{eqnarray} 
For $n$ quaternions, 
$r$ moment maps associated with $\G$ under a 
charge matrix $Q^a_i$ will read 
\begin{eqnarray}\label{}
\sum^n_{i=1} Q^a_i ~\cP_i^\dagger  {\boldsymbol \sigma}
\cP_i  ={\boldsymbol \xi}^a,
~~~~~~\cP_i=(q_i, \tilde q^\dagger_i)^t ,~~~~~~a=1, \dots, r 
\nonumber
\end{eqnarray}
or 
\begin{eqnarray}\label{DM2}
\sum^n_{i=1} Q^a_i \Big( 
|q_i|^2-|\tilde q_i|^2 \Big) =\xi^{3 a},~~~~~~
\sum^n_{i=1} Q^a_i ~  2q_i \tilde q_i
=\xi^{1 a} - i\xi^{2 a}.
\end{eqnarray}
Here, triplets ${\boldsymbol \xi}$'s are given 
level sets. By definition $\mathbb{H}^n///\G$ 
has real $4(n-r)$ dimensions.

In this note, 
we propose two kinds of GLSMs which 
have, respectively, 8D hyperK{\"a}hler manifolds and 
Calabi-Yau (CY) 4-folds as their moduli spaces. In the former 
case, we have followed pioneering works 
\cite{Harvey:2005ab, Okuyama:2005gx, 
Okuyama:2008qd}. We extend their $\N$=(4,4) 
models to 
include generic 8D hyperK{\"a}hler geometries 
dual to Type IIB $(p,q)$5-brane configurations \cite{Gauntlett:1997pk}. 
Moreover, taking an infra-red (IR) limit 
leads to frozen kinetic 
terms of vector-multiplets. 
By integrating them out, a nonlinear sigma model (NL$\sigma$M) 
can be realized 
in the Higgs branch; that is, 
the quotient 
space metric is now pulled back onto worldsheet's 
kinetic term. 
The latter case is an $\N$=(2,2) model 
which provides instead a CY 
4-fold at IR. It is not the traditional one 
\cite{Witten:1993yc} which executes a 
sympletic quotient because 
it possesses all key features of 
3D 
$\N$=2 Chern-Simons-matter theories on a stack of M2-branes 
probing toric CY 4-folds. Namely, usually 
F-term conditions (defining a master space%
\footnote{See \cite{Forcella:2008bb} for the origin of this terminology. We thanks Forcella for pointing out 
this reference.}) and 
D-term ones 
as a whole give a CY 3-fold as the story 
happens in 
4D $\N$=1 superconformal field theories (SCFT). However, because of the appearance of 
one $dualized$ $photon$ ${\cal{A}}$ here, 
the D-term constraint associated with 
${\cal{A}}$ becomes redundant 
due to 
$dynamical$ Fayet-Iliopoulos (FI) parameters and 
a 4-fold emerges thereof. 
Remarkably, we find these properties definitely 
show up in our model. We demonstrate this mechanism via one explicit example -- a 4-fold arising from $\C^3/\Z_3$.

In section 2, a Taub-NUT (ALF) space constructed by means 
of hyperK{\"a}hler quotient and its 
corresponding GLSM are reviewed. 
Then, in section 3 we write down our model mentioned above whose Higgs branch 
probes generic 8D hyperK{\"a}hler manifolds. 
Section 4 is devoted to the model probing CY 4-folds. 
Finally, our results are summarized in section 5.

\section{Review: 4D Taub-NUT (ALF) space}
To begin with, 
let us see how a multi-centered Taub-NUT 
(or ALF) space can be constructed by means of 
hyperK{\"a}hler quotient \cite{Gibbons:1996nt, Witten:2009xu}. After then, we go to 
acquaint ourselves with the GLSM description of 
it%
\footnote{For a multi-centered Taub-NUT space, the corresponding 
GLSM is considered in \cite{Okuyama:2005gx}.} according to 
\cite{Harvey:2005ab}. 
The very quotient procedure is 
exactly carried out in the Higgs branch of 
the proposed GLSM at IR. Non-zero 
hyper-multiplets take their values in 
a hyperK{\"a}hler manifold thereof. 

As said in section 1, from 
$\mathbb{H}^n \cong \mathbb{R}^{4n}$ one can construct a hyperK{\"a}hler 
manifold $\mathbb{H}^n///\G$ with $\G$ being 
triholomorphic, i.e. it does not 
alter the hyperK{\"a}hler structure. 
We will mainly adopt the following triholomorphic 
right multiplication 
\begin{eqnarray}\label{}
\Q \to \Q e^{i \theta}, ~~~~~~\theta\in(0, 4\pi] \nonumber
\end{eqnarray}
whose moment map reads 
\begin{eqnarray}\label{M}
{\boldsymbol \mu}= \Q i \bar{\Q}  =2q \tq k + 
\big( |q|^2-|\tilde q|^2 \big)i, ~~~~~~
(q, \tq )\in \C^2.  
\end{eqnarray}
In addition, a right 
multiplication $\Q p$ by a unit quaternion 
$p$ ($p\bar{p}=1$) $\in SU(2)$ does not vary 
the hyperK{\"a}hler structure $I^\alpha$ of $\Q$, while 
a left multiplication $p \Q$ which keeps the flat metric 
$ds^2=d\Q d\bar{\Q}$ invariant but rotates $I^\alpha$ plays the role of 
$SU(2)_R$ ($R$-symmetry) under 
which $(q, \tilde q^\dagger)$ transforms as a doublet. Two 
operations commute. 
From now on, we usually do not distinguish 
between ${\boldsymbol \mu}$ being pure imaginary and 
${\boldsymbol r}
=\Big(\Re(2q \tq ),   -\Im (2q \tq ), 
|q|^2-|\tilde q|^2 \Big)\in \R^3$ 
with $|{\boldsymbol r}|=|q|^2 + |\tq|^2$.

Picking up $m+1$ quaternions $\Q_a$ $(a=1,\dots,m+1)$ parameterizing 
$\mathbb{H}^{m} \times \mathbb{H} \cong \R^{4m} \times \R^4$, we want to yield a 
multi-centered Taub-NUT through 
dividing $\mathbb{H}^{m+1}$ by the following 
triholomorphic isometry ($\Q =\Q_{m+1}$):  
\begin{eqnarray}\label{}
\Q_a  \to \Q_a e^{it }, ~~~~~~ \Q \to \Q + \lambda t,
~~~~~~ \lambda \in \mathbb{R} .  \nonumber
\end{eqnarray}
There are $m$ corresponding triplet 
moment maps:
\begin{eqnarray}\label{MMMM}
{ \boldsymbol m}_a =  \Q_a i \bar{\Q}_a + 
\frac{\lambda}{2} (\Q-\bar{\Q})=  { \boldsymbol{r}}_a + \lambda 
{ \boldsymbol{y}}, ~~~~~~ \Q= y +  {\boldsymbol y}.  
\end{eqnarray}
Put into a Taub-NUT form 
the metric of $\mathbb{H}^{m} \times \mathbb{H}$ 
($r_a=|{ \boldsymbol r}_a|$): 
\begin{eqnarray}\label{me}
&&ds^2 = \sum_a d\Q_a d\bar{\Q}_a = 
\sum_a \frac{1}{4}\Big( \frac{1}{r_a} d{ \boldsymbol r}_a^2 + {r_a} 
(d\psi_a + {\boldsymbol \omega }_a d{ \boldsymbol r}_a)^2 \Big) +
dy^2 + d {\boldsymbol y}^2, \nonumber\\
&& \nabla \times {\boldsymbol \omega }_a = 
\nabla  \frac{1}{{r}_a}, ~~~~~~~~~~~~~~~~
\psi_a \in (0,4\pi].
\end{eqnarray}
By fixing the level set at 
${ \boldsymbol m}_a = { \boldsymbol \zeta}_a$ such that 
${\boldsymbol r} = -\lambda { \boldsymbol y}={ \boldsymbol r}_a - {\boldsymbol \zeta}_a$, it is straightforward to show that 
\eqref{me} becomes 
\begin{eqnarray}\label{me2}
&&ds^2 =  
U d{ \boldsymbol r}^2 + 
\sum_a \frac{|{ \boldsymbol r} + { \boldsymbol \zeta}_a |}{4}
(d\psi_a + {\boldsymbol \omega }_a d{ \boldsymbol r}_a)^2
+ dy^2, \nonumber\\
&&U_a = \frac{1}{4|{ \boldsymbol r} + { \boldsymbol \zeta}_a |} + \frac{1}{\lambda^2}, ~~~~~~~~
U=\sum_a U_a.
\end{eqnarray}
As a final step, we need to 
express 
\eqref{me2} in terms of ${\boldsymbol r}$ and a 
$gauge$-invariant $\chi$ under moment maps 
in \eqref{MMMM}, i.e.  
\begin{eqnarray}
\chi = \sum_a\psi_a -\frac{y}{\lambda}. \nonumber
\end{eqnarray}
Henceforth, instead of $\psi$'s we use variables like 
\begin{eqnarray}
(p_1, p_2, \dots, p_{m-1}), ~~~~~~~~~~
p_\alpha=\sum_{a=1}^{\alpha} \psi_a.\nonumber
\end{eqnarray}
The square $\propto (p_{\alpha}+\cdots)^2$ is completed and then dropped due to its 
$gauge$ variance. 
Finally, we are left with 
$\chi$ and $y$. 
Further taking care of the completion 
of $(y+\cdots)^2$, one arrives at 
\begin{eqnarray}
ds_{\mathrm{ALF}}^2 =  U d{ \boldsymbol r}^2 + 
\frac{1}{16} U^{-1} 
\Big( 
d\chi + (\sum_{a=1}^m { \boldsymbol \omega}_a) 
d{ \boldsymbol r} \Big)^2 \nonumber
\end{eqnarray}
where ${\boldsymbol \omega }_a$ is evaluated at 
${\boldsymbol r}+ { \boldsymbol \zeta}_a$.

\subsection{Gauged linear sigma model}

Take as the prototype the simplest one-nut $TN_4$. 
Let us review its GLSM following Harvey and 
Jensen \cite{Harvey:2005ab}. 
Introduce $\N$=(4,4) superfield conventions necessary 
for later convenience: \\
\\
$\N$=(4,4) vector-multiplet $(\Sigma, \Phi)$\\ 
($\Sigma=\frac{1}{\sqrt{2}}\bar{D}_+ D_- V$ is 
an  $\N$=(2,2) twisted chiral 
superfield)\\
$\N$=(4,4) hyper-multiplet $(Q, \tQ)$\\
$\N$=(4,4) linear hyper-multiplet $(\Psi,P)$ \\

The GLSM Lagrangian 
${\cal L}={{\cal L}_D} + {{\cal L}_F} + {\cal{L}}_{FI}$ 
for an one-nut $TN_4$ 
consists of 
\begin{eqnarray}\label{TN1}
&& {{\cal L}_D}=
\int d^4 \theta  
~\frac{1}{g^2} \Psi^\dagger \Psi
+ \frac{1}{2 g^2} \Big(P^\dagger + P +   g^2 \sqrt{2}V \Big)^2 
+ \frac{1}{e^2} \Big( -\Sigma^\dagger \Sigma  
+ \Phi^\dagger  \Phi \Big) + \Big( Q^\dagger e^{2V} Q 
+ \tilde{Q}^\dagger e^{-2V} \tilde{Q} \Big), \nonumber\\
&&{{\cal L}_F}= \int d\theta^+ d\theta^- ~
\W+ c.c. , ~~~~\W=\Big( \sqrt{2} \tilde{Q} \Phi Q -\Psi\Phi \Big),\nonumber\\
&&{\cal{L}}_{FI}=\frac{1}{\sqrt{2}}\int d\theta^+ d\bar{\theta}^- ~  t \Sigma +  \int d\theta^+ d\theta^- ~s\Phi
+c.c.  .
\end{eqnarray}
FI terms are included in ${\cal{L}}_{FI}$ 
with 
$(s, t)=(\xi^1 - i\xi^2, \xi^3 + i\theta)$ 
where superscripts are Cartesian
labels. To find out the vacuum, one has to expand 
\eqref{TN1} into (bosonic) component fields%
\footnote{Bosonic components of $\Psi$ and $P$ are as follows:  
\begin{eqnarray}
\Psi =\frac{1}{\sqrt{2}} (x^1 - ix^2 )+\cdots,  ~~~~
P= \frac{1}{\sqrt{2}}(-x^3+ i{g^2}\gamma )+\cdots, 
~~~~\gamma\sim \gamma+2\pi.
 \nonumber
\end{eqnarray}}. 
Terms from ${\cal L}_D$ in order 
are listed below (Wess-Zumino gauge of $V$) 
\begin{eqnarray}\label{D}
&&{\text {1st}}:  ~~-\frac{1}{2g^2}\Big( (\p x^1)^2 +  (\p x^2)^2 \Big) 
 \nonumber\\
&&{\text {2nd} } 
: ~~-\frac{1}{2g^2}\Big( 2 g^4(\p \gamma + A)^2  + (\p x^3)^2  \Big) -g^2 |\sigma|^2 
 \nonumber\\
&&{\text {3rd}} 
: ~~\frac{1}{e^2} (\frac{1}{2}F_{01}^2 -|\p \phi|^2-|\p \sigma|^2) 
\nonumber\\
&&{\text {4th}}
: ~~-|\D q|^2 -|\D \tq|^2 -2(|q|^2 + |\tq|^2) |\sigma|^2
\end{eqnarray}
Immediately, a D-term potential (by integrating out $D$ in $V$) 
\begin{eqnarray}\label{D1}
V_D = -\dfrac{e^2}{2}\Big( |q|^2 - |\tq|^2 -(x^3- \xi^3) \Big)^2
\end{eqnarray}
and F-term potentials (by integrating out $F_{Q}$, $F_{\tQ}$, $F_{\Phi}$ and $F_{\Psi}$)
\begin{eqnarray}\label{F1}
\begin{cases}
V_{\Phi} = -\dfrac{e^2}{2} \Big| 2q \tq -(x^1 - ix^2) 
+ s \Big|^2 \\
V_{Q}+ V_{\tQ}=  -2|q|^2 |\phi|^2-2|\tq|^2 |\phi|^2 \\
V_{\Psi}= -g^2 |\phi|^2
\end{cases}
\end{eqnarray}
follow.

Further by taking an infra-red limit ($e^2 \to \infty$) which 
decouples the vector-multiplet kinetic term, we find from 
\eqref{D} to \eqref{F1} that the $Coulomb$ branch is essentially excluded because of 
$(\phi, \sigma) \to 0$%
\footnote{There is indeed a term ${\cal L}_{top}\sim \theta F_{01}$ though 
irrelevant here.}. 
We finally obtain the supersymmetric 
vacuum 
satisfying
\begin{eqnarray}\label{sv}
|q|^2-|\tilde q|^2=x^3- \xi^3, ~~~~~~
2q \tilde q =x^1 - ix^2 -\xi^1 + i \xi^2
\end{eqnarray}
which is just the 
standard moment map considered in \eqref{M}. 
We observe that the massive gauge field $A_{\mu}$ 
(mass square $g^2$) manages to $eat$ (gauge away) one 
unwanted angular variable variant under \eqref{sv}. 
Given that the metric $|dq|^2 + |d \tilde{q}|^2$ of a 
quaternion is flat, 
we express it in a Taub-NUT form with 
$\varphi \sim \varphi+2\pi$ denoting 
the triholomorphic $U(1)$ such that 
\begin{eqnarray}\label{sv1}
&&|\D q|^2 + |\D \tilde q|^2 \Big|_{vacuum} + \frac{1}{2g^2} (\p { \boldsymbol x})^2 + \frac{g^2}{2}(\p \gamma + A)^2
= \nonumber\\
&&\Big( \frac{1}{2g^2} +  \frac{1}{4|{ \boldsymbol x}-{ \boldsymbol \xi}|} \Big) (\p { \boldsymbol x})^2
+ \frac{{|{\boldsymbol x}-{ \boldsymbol \xi}|}}{4}
(2A + 2\p\varphi + { \boldsymbol \omega}\p{ \boldsymbol x})^2 
 + \frac{g^2}{2}(\p \gamma + A)^2,\nonumber\\
&&\nabla \times{ \boldsymbol \omega} =-\nabla 
\frac{1}{| { \boldsymbol x}- { \boldsymbol \xi}|}.
\end{eqnarray}
Defining 
\begin{eqnarray}
U= \frac{1}{g^2} +  \frac{1}{2|{ \boldsymbol x}-{ \boldsymbol \xi}|}, ~~~~~~\chi=2(\gamma-\varphi),~~~~~~
\p Y=-\frac{1}{\sqrt{2}}g(\p \gamma + A), \nonumber
\end{eqnarray}
one obtains 
\begin{eqnarray}\label{sv2}
{\cal L}_{IR}= -\Big( \frac{1}{2g^2} +  \frac{1}{4|{ \boldsymbol x}-{ \boldsymbol \xi}|} \Big) (\p { \boldsymbol x})^2
- \frac{{|{\boldsymbol x}-{ \boldsymbol \xi}|}}{4}
\Big( \p\chi - { \boldsymbol \omega}\p{ \boldsymbol x} 
+2\sqrt{2}\frac{\p Y}{g} \Big)^2 
 - (\p Y)^2.\nonumber
\end{eqnarray}
Further by completing $(\p Y+\cdots)^2$,  
\begin{eqnarray}\label{sv3}
&&{\cal L}_{IR}=-\frac{1}{2} U (\p { \boldsymbol x})^2
- \frac{1}{2} U^{-1}
\Big( \frac{1}{2}\p\chi - \frac{1}{2}{\boldsymbol \omega}\p{ \boldsymbol x}  \Big)^2 .
\end{eqnarray}
Note that a remnant $\propto (\p Y + \cdots)^2$ is 
gauged away by $A_{\mu}$. To conclude, we have 
found that at IR 
($e^2 \to \infty$) the Higgs branch vacuum manifold 
manifests itself as a 
hyperK{\"a}hler quotient space. 
Integrating out $A_{\mu}$ then 
results in a NL$\sigma$M \eqref{sv2} with an 
explicit $TN_{1}$ target 
metric.

\section{8D toric hyperK{\"a}hler manifold}
We are mainly interested in 
8D toric 
hyperK{\"a}hler manifolds 
$\M_8={\mathbb{H}}^{n}/// \G$ which can be obtained by 
hyperK{\"a}hler quotient of quaternions. 
While $\M_8$=Cone(${{\cal B}}_7$) is expressed 
as a cone, the base seven-manifold 
${{\cal B}}_7$ is tri-Sasakian  
and Einsteinian with constant sectional 
curvature $R_{\mu \nu}=6g_{\mu \nu}$. 
Recently, this kind of geometry has been studied quite 
intensively in the context of AdS/CFT because 11D 
supergravity solutions of the type 
$AdS_4 \times {{\cal B}}_7$ are dual to various 
3D $\N$=3 SCFTs newly found 
in \cite{Imamura:2008nn,Jafferis:2008qz,1,2,3,Fujita:2009xz}%
\footnote{A subfamily of 
$\M_8$ called Eschenburg space ${\mathbb{H}}^{3}/// U(1)$ 
(up to an orbifold identification) 
is shown to be dual to new $\N$=3 SCFTs 
in \cite{Fujita:2009xz}.}.

As mentioned before, $\G$ is part of 
the triholomorphic isometry of 
${\mathbb{H}}^{n}$ and preserves its 
hyperK{\"a}hler structure. With respect to the remaining triholomorphic $T^2$, 
the metric of $\M_8$ can be put into the following form 
($\varphi_i\in (0, 4\pi]$): 
\begin{eqnarray}\label{MM}
\begin{cases}
ds^2 = \frac{1}{2} U_{ij} 
d {\boldsymbol{x}}_i \cdot  d {\boldsymbol{x}}_j + \frac{1}{2}U^{ij} (d\varphi_i + A_i)(d\varphi_j +A_j )\\
A_i=d{\boldsymbol{x}}_j \cdot \boldsymbol{\omega}_{ji}
=dx^a_j ~ \omega_{ji}^a, \quad \partial_{x^a_j}\omega^b_{ki}-\partial_{x^b_k} \omega_{ji}^a=\epsilon^{abc}\partial_{x^c_j} U_{ki}
\end{cases}
\end{eqnarray}
where $i,j,k=1,2$ and $a,b,c=1,2,3$ (Cartesian label)%
\footnote{Note that $U^{ij}$ is the matrix inverse of $U=U_{ij}$. }. 
Conventionally, $\M_8$ is embedded in M-theory and 
occupies (345678910), i.e. ${\boldsymbol{x}}_1=(345)$ and 
${\boldsymbol{x}}_2=(789)$ while circles 
$(\varphi_1, \varphi_2)$ stands for $(x^6, x^{10})$. 
In fact, this metric is so rigid in the sense that it is fully  determined once a proper 2 by 2 symmetric matrix $U$ 
gets specified \cite{Gauntlett:1997pk}. Certainly, adding a 
constant part $U_{\infty}$ to $U$ is still 
a solution. Issues about $U_{\infty}$ will be addressed below.

In generic non-degenerate cases where the holonomy of $\M_8$ is exactly $sp(2)$ (instead of $sp(1)\times sp(1)$), 
a fraction $3/16$ out of full 32 SUSY remains and 
there exist three real covariantly constant spinors 
rotated by $SO(3)_R$ $R$-symmetry of $\M_8$. 
Since covariantly constant spinors and the triholomorphic isometry commute, 
the same amount of SUSY survive the duality map which utilizes the above $T^2$, say, ``$\big($M-theory/two-torus$\big)$ $\leftrightarrow$ Type IIB string theory". 
Consequently, via the duality chain 
$\M_8$ gets dual to properly-oriented IIB $(p,q)$5-branes attached on a circle $\tilde{x}^6$ (T-dualized $x^6$). 
How a reduced amount of holonomy 
from $sp(2)$ to $sp(1) \times sp(1)$ which results in 
totally $1/4$ SUSY occurs? This 
becomes possible when $\M_8$ just reduces to two 
orthogonal Taub-NUT spaces occupying $(3456)$ and $(78910)$, respectively (up to some orbifold identification). 
There may be two kind of situations. 
One is when $U=U_{\infty} + \Delta U$ happens to factorize into a diagonal 
form. The other lies in zooming in on the $near$-$horizon$ 
region (${\boldsymbol x}_{1,2} \sim 0$) such that $U_{\infty}$ is gotten rid of and $\Delta U$ 
may get properly diagonalized.

As advertised, some comments follow from the relation 
\begin{eqnarray}\label{c}
\dfrac{U^{-1}}{\sqrt{\det U^{-1}}}=\dfrac{1}{\Im \tau} 
\left(
\begin{array}{cc}
   |\tau|^2 & \Re \tau  \\
\Re \tau & 1
\end{array}
\right)
\end{eqnarray}
where we have dropped the subscript $\infty$ for asymptotic 
$\tau$ and $U$ at infinity. 
Though \eqref{c} 
simply results from translating the asymptotic metric of $T^2$ in 
\eqref{MM} into its complex moduli $\tau$, this 
correlation does imply an 
$SL(2, \Z)$ covariance of both 5-brane charge vectors 
and the IIB axio-dilaton $\chi + ie^{-\phi}=\tau_{\infty}$%
\footnote{As is well-known, $\chi + ie^{-\phi}$ gets identified 
with the M-theory torus's 
complex moduli $\tau_{\infty}$ 
via the aforementioned duality chain.}. 
As a matter of fact, $U_{\infty}$ plays a role of fixing 
the orientation of 5-branes and subsequently 
determines the form of 
$\Delta U$. Therefore 
as far as preserving 
six supercharges is concerned, $U_{\infty}$ and 
$\Delta U$ are not independent. 
More precisely, the normalization is that 
when $U_{\infty}={\mathbf{1}}_{2\times 2}$ ($e^{\phi}=g_s=1$) 
a $(1,k)$5-brane%
\footnote{It occupies 
$012(3,7)_\chi(4,8)_\psi(5,9)_\theta$.} is placed 
relative to 
a (1,0)5 (NS5) brane occupying (012345) by an 
angle $\theta=\arctan k$ uniformly 
on $(3,7)$-, 
$(4,8)$- and $(5,9)$-plane. In other words, 
we have measured $\theta$ between two kinds of 5-branes 
according to \cite{Gauntlett:1997pk}
\begin{eqnarray}\label{cos}
\cos \theta = \dfrac{{\boldsymbol v}^\dagger U^{-1}_{\infty} 
{\boldsymbol v}' }
{ \sqrt{{\boldsymbol v}^2 ~{\boldsymbol v}'^2}}, 
~~~~~~~~~
{\boldsymbol v}=(p,q)^t.
\end{eqnarray}
Combining \eqref{c} and \eqref{cos}, 
we are able to tell why it is $SL(2, \Z)$ 
rather than $SL(2, \R)$ that remains as 
a symmetry of Type IIB string theory by 
supersymmetry arguments. 
Because a correct amount of SUSY should be respected in 11D 
${\mathbb R}^{1,2}\times \M_8$ due to its hyperK{\"a}hler nature, 
dual IIB $(p,q)$5-branes 
are asked to have specific orientations. 
That 
\eqref{cos} is $SL(2, \Z)$-invariant leads to 
$SL(2, \Z)$-covariant 
${\boldsymbol v}$ and $\chi + ie^{-\phi}$ 
in order to maintain definite 
3/16 SUSY necessarily.

\subsection{Gauged linear sigma model}

Equipped with the above warm-up, 
we are ready to write down a GLSM 
which provides triple moment maps from its D- and F-term 
conditions. It realizes a 8D hyperK{\"a}hler metric 
in its Higgs branch thereof at 
IR 
upon integrating out auxiliary fields. 

To begin with, 
$ {\cal L}={{{\cal L}_D}}+ {{{\cal L}_F}}$ where  
\begin{eqnarray}\label{111}
&&{{{\cal L}_D}}=
\int d^4 \theta  ~
{\boldsymbol \Psi}^\dagger {\cG} {\boldsymbol \Psi}
+ 
\frac{1}{2g_A^2} 
(P_A^\dagger + P_A + {g_A^2}\sqrt{2} {\boldsymbol p}\cdot {\boldsymbol V}  )^2 
+ 
\frac{1}{2g_B^2} 
(P_B^\dagger + P_B + {g_B^2}\sqrt{2} {\boldsymbol q}\cdot {\boldsymbol V}  )^2
 \nonumber\\
&&+ 
\sum_{i=1}^M  \Big( \frac{1}{e^2_i} ( -\Sigma_i^\dagger \Sigma_i  
+ \Phi^\dagger_i  \Phi_i ) + H_i^\dagger e^{2V_i} H_i 
+ \tilde{H}_i^\dagger e^{-2V_i} \tilde{H}_i  \Big), \nonumber\\
&&{{{\cal L}_F}}= \int d\theta^+ d\theta^-  ~\W + c.c. ,
~~~~~~
\W=
\Big( 
{\boldsymbol { H \tilde{H}}}
\cdot{\boldsymbol \Phi}
 \Big)
-\Big( \Psi_A
{\boldsymbol p} 
+ \Psi_B{\boldsymbol q}  \Big)\cdot{\boldsymbol \Phi}. 
\end{eqnarray}
This is an extension of the original work of Okuyama \cite{Okuyama:2005gx}. 
Our convention is as follows: 
\begin{eqnarray}
\label{BBBB}
&&{\boldsymbol \Psi}^\dagger=(\Psi_{A},\Psi_{B}), ~~~
\cG=\left(
\begin{array}{cc}
 \dfrac{1}{g_A^2}   & 0  \\
0 & \dfrac{1}{g_B^2}
\end{array}
\right), \\
&&\Psi_{A,B} =\frac{1}{\sqrt{2}} (x_{A,B}^1 - ix_{A,B}^2 )+\cdots,  ~~~~
P_{A,B}= \frac{1}{\sqrt{2}}(-x_{A,B}^3+ i{g^2}\gamma_{A,B} )+\cdots, 
~~~~\gamma_{A,B}\sim \gamma_{A,B}+2\pi.\nonumber
\end{eqnarray} 
Bold symbols (e.g. 
${\boldsymbol p}$, ${\boldsymbol q}$, 
${\boldsymbol \Phi}$, ${\boldsymbol V}$, 
${\boldsymbol { H \tilde{H}}}$, etc.) denote   
$M$-vectors, while $\cdot$ represents 
the usual vector inner product. Also, 
$\cG$ will be identified with the $U_{\infty}$ part in 
\eqref{MM}. 
We have turned off FI parameters which spoil the cone 
structure of $\M_8$ when one zooms in on the 
vicinity around the origin. 
In addition to couplings constants, 
$({\boldsymbol p}, {\boldsymbol q})$ are the only 
parameters in \eqref{111}. 
It becomes transparent later 
that $({\boldsymbol p}, {\boldsymbol q})$ 
is responsible for $M$ distinct charge vectors 
of IIB $(p, q)$5-branes.

We should 
expand ${\cal L}$ into components fields, 
integrate out auxiliary $D$ and 
$F$ fields. Further imposing the infra-red 
limit $e_i^2 \to \infty$  
freezes kinetic terms of vector-multiplets. 
Henceforth, we are left with (bosonic part only)   
\begin{eqnarray}\label{rest}
&&-\frac{1}{2g_A^2} (\partial {\boldsymbol x}_A)^2 - 
\frac{1}{2g_B^2} (\partial {\boldsymbol x}_B)^2 -
g_A^2 (\partial \gamma_A + {\boldsymbol p} 
\cdot{\boldsymbol A})^2 -
g_B^2 (\partial \gamma_B +{\boldsymbol q} 
\cdot{\boldsymbol A} )^2 \nonumber\\
&&+\sum_i \Big( -g_A^2 |p_i \sigma_i|^2 
-  g_B^2 |q_i \sigma_i|^2   -|\D h_i|^2 -|\D \tilde{h}_i|^2  
-2 (|h_i|^2 + |\tilde{h}_i|^2)|\sigma_i|^2
\Big)
\end{eqnarray}
and potentials  
\begin{eqnarray}\label{susy}
\begin{cases}
V_{D}= \sum_i -\dfrac{e^2_i}{2} \Big( 
|h_i|^2 -|\tilde{h}_i|^2 -(p_i x_{A}^3 + q_i x_{B}^3)
\Big)^2
\nonumber\\
V_{F} = \sum_i -\dfrac{e^2_i}{2} \Big| 
2h_i \tilde{h}_i -p_i(x_{A}^1 - ix_{A}^2)-q_i 
(x_{B}^1 - ix_{B}^2)
\Big|^2 - |\phi_i|^2 (2|h_i|^2 + 2|\tilde{h}_i|^2 + 
g_A^2 p_i^2 + g_B^2 q_i^2)\nonumber
\end{cases}
\end{eqnarray}
A SUSY vacuum accompanied by $(\phi_i, \sigma_i)\to 0$ implies 
$M$ sets of triple moment maps: 
\begin{eqnarray}\label{MH}
\begin{cases}
|h_i|^2 -|\tilde{h}_i|^2 -(p_i x_{A}^3 + q_i x_{B}^3)=0
\\
2h_i \tilde{h}_i -p_i(x_{A}^1 - ix_{A}^2)-q_i 
(x_{B}^1 - ix_{B}^2)=0
\end{cases}
\end{eqnarray}
from $V_D=V_F=0$. It is now clear that 
the model itself performs a hyperK{\"a}hler quotient 
on $M+2$ quaternions by imposing 
\eqref{MH} which kill $M$ pairs of $(h_i, \tilde h_i)$.

As before, let us rewrite the flat metric 
$|dh|^2 + |d \tilde h|^2$ of a quaternion into a Taub-NUT form 
with $\psi$ denoting the triholomorphic $U(1)$, i.e.   
\begin{eqnarray}\label{}
&&\sum_{i=1}^M    |\D h_i|^2 + |\D \tilde{h}_i|^2 \Big|_{vacuum} \nonumber\\
&&=\Delta U_{ab} 
\partial {\boldsymbol x}_a \cdot  \partial {\boldsymbol x}_b
+  \sum_{i=1}^M 
\dfrac{|{\boldsymbol X_i}|}{4}
(2A_i + 2\p \psi_i +
 {\boldsymbol \omega}_{i}\cdot
\p {\boldsymbol X}_i)^2, 
~~~~a,b =A,B \nonumber
\end{eqnarray}
where 
\begin{eqnarray}\label{}
\Delta U = \sum_{i=1}^M \dfrac{1}{4|{\boldsymbol X_i}|} 
\left(
\begin{array}{cc}
p_i^2 & p_i q_i\\
p_i q_i & q_i^2
\end{array}
\right),  ~~~~~~
{\boldsymbol X_i}= p_i {\boldsymbol x}_A + q_i 
{\boldsymbol x}_B,  ~~~~~~
\nabla \times {\boldsymbol \omega}_i = - 
\nabla\frac{1}{|{\boldsymbol X_i}|}.
 \nonumber
\end{eqnarray}
According to 
\cite{Gauntlett:1997pk}, 
$\Delta U$ corresponds to 
$M$ dual IIB $(p_i, q_i)$5-branes localized at 
${\boldsymbol X_i}=0$ w.r.t. 
$U_{\infty}=\cG$ which can be adjusted 
to $(\mathrm{const.})\textbf{1}_{2\times 2}$. 
Since 
only $\chi_{A}=\gamma_A - \sum p_i \psi_i$ and 
$\chi_{B}=\gamma_B - \sum q_i \psi_i$ 
are invariant under the action of 
moment maps in \eqref{MH}, our strategy is to let 
$M$ massive gauge fields 
gauge away remaining $gauge$-variant 
angular variables. 
To evaluate the term 
\begin{eqnarray}\label{rp}
\sum_{i=1}^M 
\dfrac{|{\boldsymbol X_i}|}{4}
(2A_i + 2\p \psi_i +
{\boldsymbol \omega}_{i} \cdot
\p {\boldsymbol X}_i)^2
-\frac{g_A^2}{2} (\partial \gamma_A + {\boldsymbol p} 
\cdot{\boldsymbol A})^2 -
\frac{g_B^2}{2} (\partial \gamma_B +{\boldsymbol q} 
\cdot{\boldsymbol A} )^2 ,
\end{eqnarray}
we adopt variables 
\begin{eqnarray}\label{}
&&\tau_k =\sum_{i=1}^k p_i {\cal{B}}_i, ~~~~~~
\tilde \tau_k =\sum_{i=1}^k q_i {\cal{B}}_i,~~~~~~
{\cal{B}}_i =A_i + \p \psi_i ,~~~~~~
k=1, \dots,M
\nonumber
\end{eqnarray}
such that every $2(A_i + \p \psi_i)$ 
splits into a sum 
\begin{eqnarray}
\frac{\tau_i-\tau_{i-1}}{p_i} + 
\frac{\tilde{\tau_i}-\tilde\tau_{i-1}}{q_i}.\nonumber
\end{eqnarray}

In dealing with \eqref{rp}, 
we complete in order  
$(\zeta_k \tau_k + \eta_k \tilde \tau_k + \cdots)^2$ 
$(\zeta_k, \eta_k=\mathrm{const.})$ from $k=1$ and gauge them away. 
Finally by substituting the outcome into 
\eqref{rest} the desired metric form as \eqref{MM} can be 
reached, i.e.  
\begin{eqnarray}\label{}
&&{\cal L}_{IR}=-U_{ab} 
\p {\boldsymbol x}_a \cdot \p  {\boldsymbol x}_b 
- \frac{1}{4}U^{ab} (\p\chi_a + \E_a)
(\p\chi_b +\E_b ),~~~~~~~ 
\chi_a \sim \chi_a +2\pi,
\nonumber\\
&&{\E}_a = -\frac{1}{2} 
\sum_{i=1}^M v_{i,a}{\boldsymbol \omega}_i \cdot 
\p {\boldsymbol X}_i ,~~~~~~~
v_{i}=(p_i, q_i). \nonumber
\end{eqnarray}
This is exactly a NL$\sigma$M which pulls back the metric 
of $\M_8$. 


\section{$\N$=(2,2) GLSM for Calabi-Yau fourfold} 

Let us propose an 
$\N$=(2,2) GLSM whose IR moduli space 
becomes a CY 4-fold. Our superfield conventions are inherited 
from section 3. The Lagrangian under consideration is 
${\cal L}={{{\cal L}_D}}+ {{{\cal L}_F}}$ where%
\footnote{We noticed incidentally during revising 
this version that there is a similar D-term in 
four dimensions considered 
in the context of ``chaotic D-term inflation" by Kawano \cite{Kawano:2007gg}. }  
\begin{eqnarray}\label{22}
&&{{{\cal L}_D}}=
\int d^4 \theta  ~ 
\frac{1}{2 g^2} 
(P^\dagger + P +  {\boldsymbol k}\cdot g^2 \sqrt{2}{\boldsymbol V})^2 
-
\sum_{a=1}^r  \frac{1}{e^2_a} \Sigma_a^\dagger \Sigma_a  
\nonumber\\
 &&+ \sum_{i}  
{\mathcal{H}}_i^\dagger ~ e^{2 V_{h(i)}} 
{\mathcal{H}}_i ~ e^{-2 V_{t(i)}}   
+ \sum^r_{a=1} \Phi_{a}^\dagger e^{2V_a} \Phi_a,\nonumber\\
&&{{{\cal L}_F}}= \int d\theta^+ d\theta^-  ~\W + c.c. ,
~~~~~~~~~~
{\boldsymbol k}=(k_1, \cdots,k_r)\in 
\Z^r.
\end{eqnarray}
Note that $\W$ denotes the superpotential and the gauge invariance in the first term of 
${\cal L}_D$ is maintained by  
\begin{eqnarray}
{\boldsymbol k}\cdot {\boldsymbol V} \to  
{\boldsymbol k}\cdot {\boldsymbol V}+ (\Lambda^\dagger +\Lambda), 
~~ ~ ~ P\to P-\sqrt{2}g^2 \Lambda, 
~~ ~ ~ P^\dagger\to P^\dagger-\sqrt{2}g^2 \Lambda^\dagger.
\nonumber
\end{eqnarray}
Bi-fundamental chiral fields ${\mathcal{H}}_i$ 
charged under $r$ $U(1)$'s 
according to a given quiver diagram with $r$ 
nodes. 
The subscript $i$ runs over arrows representing 
bi-fundamental fields 
in the quiver diagram, while $h(i)$ or 
$t(i)$ specifies a node 
on which the head or tail of $i$ ends. 
In addition, $\W$ subject to holomorphy is again 
related 
to the given quiver%
\footnote{Based on the 
$brane$ 
$tiling$ technology one can read off $\W$ from 
a given quiver. See \cite{Yamazaki:2008bt} for an excellent review.}. 
Note that only $P$ 
of the previous $\N$=(4,4) multiplet $(\Psi, P)$ enters 
${\cal L}$ with the 
lowest component $x+i\gamma$ ($\gamma\sim\gamma+2\pi$). 
It is apparent that both ${\cal L}$ and 
the resultant IR moduli space are characterized only 
by ${\boldsymbol k}$ and the responsible quiver diagram.

First of all, let us expand superfields in ${\cal L}$ into components and integrate 
out auxiliary $D$ and $F$ fields. Taking $(e_a^2, g^2) \to \infty$ freezes kinetic terms of fields 
$(A_a, \sigma_a,  x)$. 
What we are left with are (bosonic part only)
\\
1. Kinetic 
terms of $(h_i, \phi_a)$ 
(bosonic component of $({\mathcal{H}}_i, \Phi_a)$ \\
2. A remnant: 
\begin{eqnarray}\label{dua}
-g^2 (\p \gamma + {\boldsymbol k}\cdot {\boldsymbol A} )^2
\end{eqnarray}
3. A potential:
\begin{eqnarray}\label{SV}
&&V_{pot}=\sum_{i}|{{h}}_i|^2 
| Q^{h(i)}_i \sigma_{h(i)} 
-  Q^{t(i)}_i \sigma_{t(i)}|^2 
+ \sum_a 2|\sigma_a|^2 |\phi_a|^2
\nonumber\\
&&+\sum_a g^2 |k_a \sigma_a|^2  + \sum_i  
\Big| \frac{\p \W_{}}{\p h_i} \Big|^2 + 
\sum_a \Big| \frac{\p \W}{\p \phi_a} \Big|^2
+V_D, \nonumber
\end{eqnarray}
where 
\begin{eqnarray}\label{DDD}
V_D= \sum_a \frac{e^2_a}{2} \Big( 
\sum_i (\delta_{h(i)}^a Q^a_i   - \delta_{t(i) }^a 
Q^{a}_i )|h_i|^2  + |\phi_a|^2
-k_a x 
\Big)^2 = \sum_a \frac{e^2_a}{2} D_a^2 
\end{eqnarray}

The vacuum manifold $\M$ 
is determined by 
\begin{eqnarray}\label{}
\sigma_a=\phi_a=0~~~~~~D_a=0, ~~~~~~{d \W_{}}=0.
\end{eqnarray}
Several comments follow. 
A distinguishing feature departing 
from those dealt with in previous sections is that 
by means of $\gamma$'s 
e.o.m. one gauge field ${\cal A}={\boldsymbol k}\cdot {\boldsymbol A}$ in \eqref{dua} can be 
dualized into a scalar $-\gamma$. 
In the context of 3D Yang-Mills-Chern-Simons theory, ${\cal A}$ is referred to as a $dual$ $photon$. 
Together with $V_D=0$ we can 
understand that 
the model itself executes rather a $sympletic$ quotient 
over complex variables $h_i$ than a hyperK{\"a}hler one. 

Let us analyze the moduli space in detail 
from both field-theoretical and geometrical 
grounds. 
\\
1. The constraint $\sum_a D_a=0$ in \eqref{DDD} characterizes the decoupled 
diagonal $U(1)$ and implies as well that $\sum_a k_a=0$ 
or the vector 
${\boldsymbol k}$ is orthogonal to $(1,\cdots,1)$. To this end, 
one has only $(r-2)$ linearly independent 
D-term conditions \cite{Martelli:2008si}: 
\begin{eqnarray}\label{}
\sum_a \ell_a  
D_a =0, ~~~~~~~~~~~~
{\boldsymbol \ell} \in 
\mathrm{Ker}({\boldsymbol k}), ~~~~~~~~~~~~{\boldsymbol \ell}\ne (1,\cdots,1). \nonumber
\end{eqnarray} 
This fact is consistent with that we have a 
dualized $\cal{A}$ and leads naturally to $\M=\{ d\W=0 \}//C^{\star r-2}$ which 
is a CY 4-fold by definition emerging from a 3-fold. 
Namely, $\M=\{ d\W=0 \}//C^{\star r-1}$ is a CY 3-fold because its derivation 
is just the same with that of 4D $\N$=1 SCFTs on 
a bunch of D3-branes probing CY$_3$ cones. \\
\\
2. Geometrically, 
an ungauged $U(1)$ ${\cal{A}}$ suggests the existence of 
an $S^1$ in the vacuum 
moduli space correspondingly. 
According to $V_D=0$, we can thus describe $\M$ as 
a 3-fold fibered on a real line $x$ with 
a circle bundle fibered over them with non-trivial 
first Chern class \cite{Aganagic:2001ug, Aganagic:2009zk}. 
We find that the field 
$x$ in \eqref{DDD} (or $V_D=0$) serves as $dynamical$ FI 
parameters and plays the very role of giving varying K{\"a}hler classes for the fibered 3-fold. \\
\\
3. Let us elaborate on arguments about the $surviving$ $S^1$ bundle associated with ${\cal{A}}$. For conceptual convenience, we can first think of 
it as $\gamma$ charged logarithmically under 
$r$ $U(1)$'s: 
\begin{eqnarray}\label{GGG}
\gamma \to \gamma+\sum_{a=1}^r k_a \lambda_a, ~~~~~~~~
\gamma \sim \gamma+2\pi.
\end{eqnarray}
Consequently, \eqref{GGG} defines a one-parameter 
gauge transformation ${\boldsymbol \lambda} 
=t(k_1,\cdots, k_r)$. We find that $\gamma$ 
is especially helpful when one wants to determine the correct periodicity of $S^1$. 
In fact, there may remain some 
discrete amount of 
gauge symmetry $\Gamma$ of $\cal{A}$ such that 
$\Gamma$ shrinks the circumference of $S^1$ to 
$2\pi/gcd({\boldsymbol k})$. This is apparent from 
the expression of 
${\boldsymbol \lambda}$. Let us put things in an 
inverse way. From \eqref{dua} $\cal{A}$ is spontaneously broken 
once $\gamma$ acquires a vev $\gamma_0$. But if there exists $f=gcd({\boldsymbol k})$ with $\gamma_0=f\gamma'_0 + {2m\pi}$ $(m\in\mathbb{Z})$, the gauge 
fixing becomes incomplete due to a factor ${2m\pi}/{f}$. This soon means 
that the gauge symmetry of $\cal{A}$ is only 
spontaneously broken down to a discrete extent. 
To conclude, we should at the end divide $\M$ further 
by $\Gamma$. All these 
are directly reminiscent of the novel 
mechanism in newly-developed 3D 
$\N$=2 Chern-Simons-matter 
theories \cite{Aharony:2008ug}. 
Therefore, 
it turns out that, 
up to a discrete quotient $\Gamma$, 
$\M=\{ d\W=0 \}//C^{\star r-2}$ where additional 
real two dimensions result 
effectively from degrees of freedom of $x+i\gamma \subset P$%
\footnote{Typically, $x$ and $\gamma$ are called Fayet-Iliopoulos and 
Stuckelburg fields, respectively.}.

\subsection{Toric geometry}
Let us review quickly stuffs about toric varieties 
which are highly helpful when one lifts 
a toric 
CY$_3$ to a toric CY$_4$. In general, a toric variety $V$ 
is expressed as 
$\C^n \backslash F//\C^{\star \cG}$ 
and can be summarized pictorially by a toric diagram%
\footnote{The subset $F$ is determined by partial resolutions.} 
consisting of $n$ $d$-vectors $\nu_i\in \Z^d$ 
$(d=n-\cG)$ subject to  
\begin{eqnarray}\label{}
\sum_{i=1}^n ~Q^A_i ~\nu_i =0,~~~~~~A=1, \dots, \cG,  ~~ ~~ ~~ 
Q^A_i: \mathrm{charge ~matrix}.
\end{eqnarray}
By imposing the Calabi-Yau condition: 
$\sum_i Q^A_i=0$~ $\forall A$, these 
$\nu$'s can be written as $\nu_i =(1, \cdots)^t$ and 
are called toric data of a toric 
CY $d$-fold. For $d$=3, toric data are plotted on a 
plane, while $d$=4 toric data are encoded 
in 3D lattice points which define a convex polytope -- 
$crystal$. This is why we adopt the name $crystal$ in our title page because 
CY 4-folds are under consideration.

Let us talk about 
the geometric meaning of toric data. Every 
$\nu_i$ assigns a shrinking 1-cycle out of $T^d$ 
at 
the $i$-th facet of the boundary 
$\p {\cal{C}}_+$ where 
${\cal{C}}_+$ denotes a cone in $\Z^d$ over which 
$T^d$ is fibered over. In $d$=3 cases, 
lines where two facets meet will altogether constitute a $web$ diagram 
which in turn represents a tree formed by multiple 
semi-infinite NS5-branes. 
Although we will not 
go further to details, interested readers are 
encouraged to consult \cite{Yamazaki:2008bt}.

\subsection{Calabi-Yau crystal}
Though lifting 3-folds to 4-folds has 
been studied quite heavily to date%
\footnote{See also 
\cite{Ueda:2008hx,Imamura:2008qs,Taki:2009wf} 
and references therein.}, let us just 
pick up one canonical example $\C^3/\Z_3$ 
discussed also in 
\cite{Martelli:2008si, Hanany:2008cd, Aganagic:2009zk}. 
Here, $\Z_3$ is embedded in the Hopf $U(1)$ bundle of $\C^3$ whose action is  
\begin{eqnarray}\label{}
(z_1, z_2,z_3)\to (\epsilon z_1, \epsilon z_2,\epsilon z_3), 
 ~~~~~~~~~ \epsilon=e^{2\pi i /3}. 
\end{eqnarray}

\subsubsection{$\C^3/\Z_3$}
A stack of D3-branes probing $\C^3/\Z_3$ 
\cite{Douglas:1996sw, Douglas:1997de} has 
nine bi-fundamental chiral superfields on its 
worldvolume gauge theory with a 
superpotential (though we will focus 
on its abelian version) 
\begin{eqnarray}\label{}
\W\propto \epsilon^{IJK} \Tr (X_I Y_J Z_k) ,~~~~~~~~~ 
I,J,K=1,2,3. 
\end{eqnarray}

\begin{figure}[t]
   \begin{center}
     \includegraphics[height=8cm]{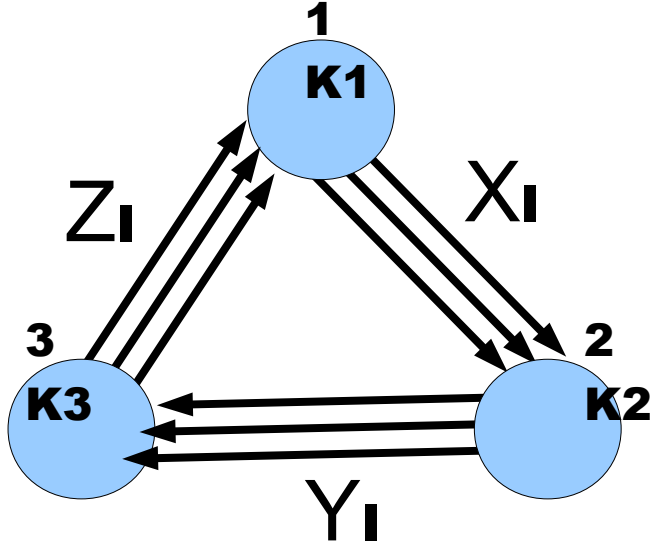}
   \end{center}
\caption{ A quiver diagram for $\C^3/\Z_3$. 
Three nodes are labeled by integers 
$(k_1, k_2,k_3)$. 
Three sets of chiral superfields are denoted as 
$(X_I, Y_I, Z_I)$ $(I=1,2,3)$. }
\label{FFF1}
\end{figure}

In Figure \ref{FFF1}, three nodes denote 
three gauge groups and hence $(3-1)$ D-term conditions. 
In addition, $d\W=0$ renders four more constraints 
\begin{eqnarray}\label{ft}
X_1 Y_2 =X_{2} Y_{1}, ~~~~~~
X_2 Y_3 =X_{3} Y_{2}, ~~~~~~(X,Y) \leftrightarrow (Y,Z).
\end{eqnarray}
Instead, we can $solve$ \eqref{ft} via 
$X_I = u_1 v_I$, $Y_I = u_2 v_I$ and 
$Z_I = u_3 v_I$ $(I=1,2,3)$ accompanied by a D-term like condition: 
\begin{eqnarray}
|u_1|^2 +|u_2|^2 
+|u_3|^2 -|v_1|^2 -|v_2|^2 -|v_3|^2=0.
\end{eqnarray}
$\M_F =\{ d\W=0 \}$ can then 
be thought of as $\C^6//\C^{\star }_{(1,1,1,-1,-1,-1)}$. 
Alternatively, the following exact sequence 
\begin{eqnarray}\label{es}
0 \to \Z^{} \stackrel{ Q^{t}}{\to} \Z^{6} 
\stackrel{T}{\to} \mathbb{N} \to 0
\end{eqnarray}
reveals the same thing. 
It merely says mathematically 
$ \M_F$ can be regarded as 
$\C^{6}//\C^{\star }$ 
where the action of $\C^{\star }$ is spelt out by 
$Q$. 
More precisely, one can express $\M_F$ as a cone $\mathbb{M}_+$ in $\mathbb{M}=\Z^{6}$. Its dual cone (toric data) 
$\mathbb{N}_+$ in $\mathbb{N}=\mathrm{Hom}(\mathbb{M},\Z)$ is generated by six lattice points, i.e. 
$T$: $\Z^{6} \to \mathbb{N}$. 
By definition Ker($T$) is the image of $Q^t$, so 
the cokernel of $Q^t$ is isomorphic to $T$ 
($Q$: 1 by 6 matrix). To get a 3-fold, 
we need to incorporate D-term constraints as well. 
In summary, we can obey three steps below \cite{Douglas:1997de}: 
\\
1. Introduce $U$ (5 by 6 matrix) with $UT^{t}=1$ and $V$ 
(2 by 5 matrix). $V$ encodes two D-term constraints on 
five independent variables as indicated in \eqref{ft}  \\
2. Concatenate 
$VU$ (2 by 6 matrix) and $Q$ to yield $Q'$ 
(3 by 6 matrix) \\
3. Computing the cokernel of $Q'^t$ 
gives us desired toric data in terms of a map 
${\cal{T}}$: $\Z^6 \to \Z^3$\\

To have a 4-fold we need to use a new $V$ involving only 
one linearly-combined D-term constraint 
(irrelevant to ${\cal{A}}$) out of three nodes and repeat the 
above procedure. 
Instead of this approach, 
a more convenient one figured out 
in \cite{Martelli:2008si} 
is to directly perform one more 
quotient on previous 
$(u_I, v_I)$ by another $\C^{\star}$: 
$(k_1+ k_2,-k_1, -k_2,0,0,0)$. 
Translating this into the action 
on $(X_I, Y_I)$, one easily realizes that it 
is the combination $-k_2 Q_1 + k_1 Q_2$ orthogonal to 
${\boldsymbol k}\cdot {\boldsymbol Q}$ 
with ${\boldsymbol k}=(k_1, k_2, k_3=-k_1-k_2)$. 
See Table 1 for informations about 
$Q_i$ responsible for the $i$-th node. 
\begin{table}[h]
\caption{Nine bi-fundamental 
chiral matters charged under 
three gauge groups in the case of ${\mathbb{C}}^3/{\mathbb{Z}}_3$}
  \begin{center}
    \begin{tabular}{cccccccccc}\hline
   &$X_1$&$X_2$&$X_3$&$Y_1$&$Y_2$&$Y_3$&$Z_1$&$Z_2$&$Z_3$ \\ \hline
$Q_1$& -1&    -1&   -1&    0&   0&    0&    1&    1&1\\ \hline
$Q_2$&1&    1&   1&    -1&   -1&    -1&    0&    0&0\\ \hline
$Q_3$&0&    0&   0&    1&   1&    1&    -1&   -1&-1 \\  \hline
    \end{tabular}
  \end{center}
\end{table}
Arbitrary 
${\boldsymbol k}$ encountered in \eqref{22} 
meet the requirement $\sum_a k_a=0$. Also, we may demand $gcd(k_1, k_2)=1$ 
technically in order to 
avoid a further discrete quotient $\Gamma$ explained around \eqref{GGG}.

All in all, via $(k_1, k_2,)=(2p-k, k-p)$ 
we end up with an eight cone over a seven Sasaki-Einstein 
base denoted as $Y^{p,k} ({\mathbb{CP}}^2)$. Its 
toric data are \cite{Martelli:2008si}: 
\begin{eqnarray}\label{}
&&w_1=(0,0,0), ~~
w_2=(0,0,p),~~
w_3=(1,0,0),\nonumber\\
&&w_4=(0,1,0),~~
w_5=(-1,-1,k),~~
w_0=(0,0,k-p).
\end{eqnarray}
These six vectors do form a 3D crystal as was shown in 
\cite{Martelli:2008rt} with an additional vector $w_0$ standing for 
a resolution, provided $p\le k\le 2p$. 
See \cite{Martelli:2008si} for more details.

\section{Summary}
In this short note we have studied two kinds of GLSMs whose 
vacuum moduli spaces are real eight-dimensional 
hyperK{\"a}hler and Calabi-Yau manifolds, respectively. 

The Higgs branch of the 
former $\N$=(4,4) GLSM manifests itself as 
a generic 8D hyperK{\"a}hler quotient space at the 
infra-red regime. 
On the other hand, the latter $\N$=(2,2) 
GLSM has a more flexible 
superpotential $\W$ as well as chiral matter contents 
spelt out by a quiver diagram. 
Usually, D- and F-term constraints 
give rise to a toric CY 3-fold 
just as in familiar 4D $\N$=1 SCFT setups. Nevertheless, 
owing to a novel mechanism introduced 
in section 4, 
3-folds are promoted to 4-folds exactly the same way 
as what happens in recently-developed 
3D $\N$=2 Chern-Simons-matter theories which have 
toric CY 4-folds as their moduli spaces. 
In words, 
in addition to the diagonal gauge field $\sum_a A_a$, an 
ungauged dual photon 
${\cal{A}}=\sum_a k_a A_a$ as well does not enter 
the quotient process due to $dynamical$ 
Fayet-Iliopoulos parameters. Therefore, 
the moduli space dimension is enhanced from six to 
eight. The extra 
two dimensions effectively result from 
degrees of freedom of a chiral superfield 
$P$. Parameters $k_a$ in our 2D Lagrangian satisfying $\sum_a k_a=0$ 
are arbitrary integers. Their role is 
analogous to 3D quiver Chern-Simons levels. 
We discussed the whole story  
via a very well-developed 3-fold $\C^3/\Z_3$.

\section*{Acknowledgements}
We would like to thank Amihay Hanany, Kazuo Hosomichi, Hiroaki Kanno, Muneto Nitta, 
Hiroshi Ohki, Yutaka Sakamura and especially 
Kazumi Okuyama for enlightening discussions at different intermediate stages. 
TST is grateful to 
organizers of the wonderful workshop on ``Branes, Strings and Black Holes" at Yukawa Institute of Theoretical Physics. 
TST is supported in part by the postdoctoral program
at RIKEN.

\end{document}